\documentclass[prl,aps,superscriptaddress,amsmath,amssymb,showpacs,twocolumn,floatfix]{revtex4}
\usepackage{hyperref}
\usepackage{epsfig}
\newcommand{\Hef}{\mathcal{H}_\textrm{eff}}
\newcommand{\ket}[1] {\left|#1\right\rangle}
\newcommand{\Ham}{\mathcal{H}}
\newcommand{\rem}[1]{}
\begin{document}

\title{Crossover from adiabatic to antiadiabatic quantum pumping with dissipation
}

\author{Franco Pellegrini}
\affiliation{SISSA, Via Bonomea 265, I-34136 Trieste, Italy}
\affiliation{CNR-IOM Democritos National Simulation Center, Via Bonomea 265, I-34136 Trieste, Italy}

\author{C. Negri}
\affiliation{Laboratoire d'Ondes et Mati\`ere d'Aquitaine, (UMR 5798), CNRS and Universit\'e de Bordeaux I,
351 Cours de la Lib\'eration, F-33405 Talence Cedex, France}

\author{F. Pistolesi}
\affiliation{Laboratoire d'Ondes et Mati\`ere d'Aquitaine, (UMR 5798), CNRS and Universit\'e de Bordeaux I,
351 Cours de la Lib\'eration, F-33405 Talence Cedex, France}
\affiliation{Laboratoire de Physique et Mod\'elisation
des Milieux Condens\'es, (UMR 5493), CNRS and Universit\'e Joseph Fourier,
Grenoble F 38042, Cedex, France}

\author{Nicola Manini}
\affiliation{SISSA, Via Bonomea 265, I-34136 Trieste, Italy}
\affiliation{CNR-IOM Democritos National Simulation Center, Via Bonomea 265, I-34136 Trieste, Italy}
\affiliation{ETSF and Dipartimento di Fisica, Universit\`a degli Studi di Milano, Via Celoria 16, 20133 Milano, Italy}

\author{Giuseppe E. Santoro}
\affiliation{SISSA, Via Bonomea 265, I-34136 Trieste, Italy}
\affiliation{CNR-IOM Democritos National Simulation Center, Via Bonomea 265, I-34136 Trieste, Italy}
\affiliation{International Centre for Theoretical Physics (ICTP), P.O. Box 586, I-34014 Trieste, Italy}

\author{Erio Tosatti}
\affiliation{SISSA, Via Bonomea 265, I-34136 Trieste, Italy}
\affiliation{CNR-IOM Democritos National Simulation Center, Via Bonomea 265, I-34136 Trieste, Italy}
\affiliation{International Centre for Theoretical Physics (ICTP), P.O. Box 586, I-34014 Trieste, Italy}

\date{\today}

\begin{abstract}
Quantum pumping, in its different forms, is attracting attention from
different fields,  from fundamental quantum mechanics, to nanotechnology, to 
superconductivity.
We investigate the crossover of quantum pumping from the adiabatic to the
anti-adiabatic regime in the presence of dissipation, and  find 
general and explicit 
analytical expressions for the pumped current in a minimal model
describing a system with the topology of a ring 
forced by a periodic modulation of frequency $\omega$.
The solution
allows following in a transparent way the evolution of 
pumped DC current
from much smaller to
much larger  $\omega$ values 
than the other
relevant energy scale,
the energy splitting introduced by the modulation.
We find and characterize 
a temperature-dependent optimal value of the
frequency for which the pumped current is maximal. 
\end{abstract}

% 85.25.Cp : SQUIDS?  (NICK)

\maketitle

A current with a net DC component can be pumped in an electronic system 
without leads and bias voltages, through a ``peristaltic'' 
modulation of the transmission amplitudes and gate voltages \cite{pumping}.
This effect has both classical and quantum components \cite{Brouwer98,Cohen03}, 
and occurs for unpaired electrons as well as for Cooper pairs \cite{Pekola10,Russomanno10}.
When the modulation is adiabatic i.e., when the pumping period is much
longer than the intrinsic time-scale of the system, so that transitions
between states do not occur, it has long been recognized that the charge
pumped over a period has a geometric nature
\cite{Brouwer98,Cohen03,MottonenPRB06} and is in many cases quantized.
These geometrical aspects survive even in the presence of a coupling
between the electrons and an external phonon bath, despite the obvious
source of inelastic effects represented by the bath
\cite{Fioretto_PRL08,Pekola10,Russomanno10}.
Unsurprisingly, like in classical pumps, the current in this slow,
adiabatic regime increases proportionally to the driving frequency
$\omega/2\pi$, as long as $\hbar\omega$ is much smaller than all intrinsic
energy scales of the system.

The question we address in this Letter is: of what kind, and of what
magnitude are the deviations from adiabatic pumping that will show up in
the DC current when the pumping frequency grows higher and higher?
What is the behavior of the DC current as frequency crosses over beyond
the adiabatic and into the antiadiabatic ($\omega\to \infty$) regime?
To obtain specific answers, we shall focus on the crossover from adiabatic to antiadiabatic 
quantum pumping (or ``stirring'' \cite{Cohen03,Hiller08,Salmilehto10}) in a system 
with the topology of a ring, in the presence of dissipation.
For a particular but reasonable choice of coupling to the bath, we find
that the dissipative model admits a full
analytical solution for the steady state current valid at arbitrary frequency.
%
%SCUSATE LA POMPA, MA MI PARREBBE UTILE "CARICARE" UN PO', SE NO IL MODELLO
%E' DAVVERO SOLO UN GIOCATTOLO...(ERIO)
%
Through that solution we can analyze and understand the main predicted
features of pumping-frequency dependence of the DC current.
At low frequencies the pumped current tracks the known adiabatic result,
namely DC current increases linearly with frequency, and the pumped charge
is as expected geometric in nature (albeit not quantized).
However, and this is a surprising outcome, the pumped DC current turns
nonmonotonic for increasing $\omega$, going through a temperature-dependent 
%maximum  % AGGIUSTAMENTO
optimal value
and then dropping %to vanish 
eventually as $\omega^{-1}$ for $\omega\to \infty$.
%
%Moreover, and this is the second surprise, the DC current at the maximum
%generally {\it exceeds} the current pumped in the absence of dissipation,
%and that by a temperature dependent factor reaching the limiting value of 2
%as the bath temperature vanishes.
%SIETE D'ACCORDO CHE ANCHE QUESTO SECONDO FATTO SOPRA E' SORPRENDENTE? PER
%ME LO E' STATO...(ERIO)
%PURTROPPO.... (Nick)
%
Beyond the strict limits of the present model, we also surmise that these
results are representative of a larger class of orbital doublet systems
weakly coupled to a generic environment.
%

%%%%%%%%%%%%%%%%%%%%%%%%%%%%%%%%%%%%%%%%%%%%%%%%%%%%%%%%%%%%%%%%%%%%%%%%%%%%%%
\begin{figure}\centering
\includegraphics[width=.25\textwidth]{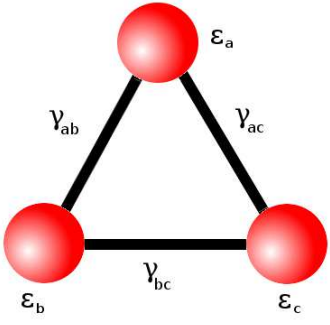}
\caption{\label{Model:fig} 
Schematic of the minimal ring  model,
realized by three quantum-dots $a,b,c$  with externally controlled potentials 
$\epsilon_a$, $\epsilon_b$ and $\epsilon_c$ used for pumping.
The $\gamma_{ij}$ are here inter-dot hoppings, generally fixed.
An alternative realization could be a molecular trimer
\cite{%Gerber78,
Delacretaz,Dugourd90} % EVENTUALMENTE RIMUOVERE X SCORCIARE
where three atomic orbitals $\ket{a}$, $\ket{b}$ and $\ket{c}$ 
have identical energies, while the hopping integrals 
$\gamma_{ij}$ can actuate the pumping by modulating  atomic displacements.
In either case the relevant states of the system
can be mapped to the pseudospin-$1/2$ model of Eq.~\eqref{HS}.
}
\end{figure}
%%%%%%%%%%%%%%%%%%%%%%%%%%%%%%%%%%%%%%%%%%%%%%%%%%%%%%%%%%%%%%%%%%%%%%%%%%%%%%%

Consider the minimal model constituted by a three-site ring as in
Fig.~\ref{Model:fig}, each of the identical sites $a,b,c$ endowed with a
single nondegenerate electronic level of energy $\epsilon_{i}(t)$, and
different sites coupled by hoppings $\gamma_{ij}(t)$, where $i,j=a,b,c$.
Current pumping can be obtained, for instance, by letting
$\gamma_{ij}=\gamma_0$ and externally actuating a cyclic variation of the
three on-site energies
$\epsilon_{i}(t)=-\hbar\Delta\cos{(\omega t + \phi_i)}$, with
$\phi_a=0$, $\phi_b=-2\pi/3$, $\phi_c=+2\pi/3$.
In the perturbative limit ($\hbar\Delta \ll \gamma_0$) the three-site ring
can be replaced by a simpler effective orbital pseudospin model obtained by
removing the totally symmetric state
$\ket{0}=\left(\ket{a}+\ket{b}+\ket{c}\right)/\sqrt{3}$  (of energy
$-2\gamma_0$ for $\Delta=0$, doubly occupied and irrelevant)
to retain only the two states $\ket{x}=(\ket{b}-\ket{c})/\sqrt{2}$,
and $\ket{y}=(2\ket{a}-\ket{b}-\ket{c})/\sqrt{6}$,  orbitally degenerate
in the unperturbed ring $\Delta=0$, with energy $\gamma_0$.
% NICK: NON CAPISCO L'UTILITA` di QUESTO COMMENTO A QUESTO PUNTO:
% ( we generally set <$\epsilon_{i}(t)$>=0).
% FORSE SIGNIFICA
% (we generally set $\sum_i\epsilon_i(t)=0$).
% MA ABBIAMO GIA` DATO SOPRA UNA FORMULA PER GLI \epsilon_i, SOPPRIMO
%
The single mobile electron now occupies the orbital doublet  
$\ket{x}$, $\ket{y}$, leading to a pseudospin-$1/2$ problem
with the time-dependent Hamiltonian \cite{EPAPS_current}:
\begin{equation} \label{HS}
\Ham_S(t) = \frac{\hbar\Delta}{2}
\left[\cos(\omega t)\sigma^z + \sin(\omega t)\sigma^x\right] \;,
\end{equation}
where $\sigma^{\xi}$ are Pauli matrices. 
%
%(A similar result is obtained also for a single electron in the presence of a
%half quantum of flux.) 
%QUESTO COMMENTO ERA GIUSTO MA NON MI PARE NECESSARIO, E MANGIA SPAZIO (ERIO)
%PER ME E` INDIFFERENTE LASCIARE O TOGLIERE, NON ABBIAMO PROB di SPAZIO (NICK)
%
The current
$I(t)=\langle I_{ab}\rangle =
-iq_{\rm e}\gamma_0 \langle c_b^\dagger c_a-c_a^\dagger c_b \rangle$
(where $q_{\rm e}$ is the elementary charge)
in the pseudospin representation is given by
$I(t)=I_0 \langle \sigma^y \rangle$, where $I_0=q_{\rm e}\gamma_0/\sqrt{3}$. 

While this is the same type of Hamiltonian previously used to study
adiabatic pumping \cite{Pekola10}, we now find that this problem allows a
more general analytical solution for arbitrary pumping frequency beyond the
adiabatic regime, and in the presence of a coupled bath, so long as the
coupling is weak.
%E' GIUSTO DIRE COSI'? (ERIO)
% HO TOLTO L'OHMIC: FABIO HA CHIARITO SOTTO CHE L'EQUAZIONE A COEFFICIENTI
% COSTANTI VIENE COMUNQUE ANCHE SE L'ACCOPPIAMENTO E` NON OHMICO, QUINDI
% NON E` IL CASO DI INSISTERE QUI (NICK)
%
The exact time evolution induced by (\ref{HS}) can be obtained by noting that 
$H=(\hbar\Delta/2) R_y(\omega t) \sigma_z R^{-1}_y(\omega t)$, where 
$R_y(\omega t)=e^{-i\omega t\sigma^y/2}$, represents a 
uniform rotation by an angle $\omega t$ around the $y$-axis.
Performing this time-dependent unitary transformation and defining 
$|\psi(t)\rangle = R_y(\omega t) | \tilde{\psi}(t)\rangle$,
the Schr\"odinger equation for $|\tilde{\psi}\rangle$ is governed by the
effective Hamiltonian
\begin{equation} \label{Heff}
\Ham_{\rm eff} =
   R^{-1}_y \Ham_S R_y - iR^{-1}_y \dot{R}_y =
   \frac{\hbar}{2}\left( \Delta \sigma^z - \omega\sigma^y \right)
   =\frac{\hbar\omega'}{2} \hat{\bf n} \cdot \boldsymbol{\sigma} \,.
\nonumber
\end{equation}
This now represents a time-independent field pointing in the direction  
$\hat{\bf n}=(0,-\omega/\omega',\Delta/\omega')$,
where $\omega'=\sqrt{\omega^2+\Delta^2}$ is the associated Larmor frequency.
The problem thus has a simple solution in this reference frame: the spin state
$|\tilde{\psi}\rangle$ precesses around $\hat{\bf n}$, 
while the current retains the form
$I(t)=I_0\langle \tilde{\psi}(t)| \sigma^y |\tilde{\psi}(t) \rangle$.
The current carried by the eigenstates $|\hat{\bf n};\pm \rangle$ of
$\Ham_{\rm eff}$ 
is
$I_0 \langle {\bf n};\pm | \sigma^y |\hat{\bf n};\pm\rangle =
\mp I_0 \omega/\omega'$, respectively.
In the absence of coupling to the bath, all time dependence of
the current is determined just by the initial conditions.
In particular, the two eigenvectors of $\Ham_{\rm eff}$ carry a 
pure (and opposite) DC current, while any other initial condition
yields a DC plus an AC current.
The DC component is determined by the projection of the pseudospin onto the
eigenstates of $\Ham_{\rm eff}$:
\begin{equation}\label{current}
I = I_0 \, P \, \frac{\omega}{\omega'} \;,
\end{equation}
with the pseudospin polarization
$P = -{\rm Tr}(\hat{\bf n} \cdot \boldsymbol{\sigma} \, \tilde\rho_S)$
expressed in terms of the density matrix operator $\tilde\rho_S$ in the
rotating frame.

Even if one prepares the initial density matrix in a pure state, the
slightest dissipation will eventually drive the system to a different
(generally periodically time-dependent) steady state.
To describe the effect of dissipation, we introduce the environment in the
standard form \cite{Weiss99} of a heat bath of harmonic oscillators at
temperature $T$ linearly coupled to the charge fluctuations, embodied in
this system by the two operators $\sigma_z$ and $\sigma_x$.
The dissipative part of the Hamiltonian is thus $\Ham_B + \Ham_{SB}$ where:
\begin{eqnarray}\label{HB}
\Ham_B &=&
\sum_{\xi=z,x} \sum_\nu
\left[\frac{p_{\xi,\nu}^2}{2m}+\frac{m\omega_\nu^2 q_{\xi,\nu}^2}{2}\right] \;,
\\ \label{HSB} 
\Ham_{SB} &=&
\sum_{\xi=z,x} \sum_\nu 
\sqrt{\frac{2 m\omega_\nu}{\hbar}}
\lambda_{\xi,\nu} q_{\xi,\nu} \sigma^\xi \;.
\end{eqnarray}
Here $\omega_\nu$ are the oscillator frequencies and $\lambda_{\xi,\nu}$ are 
coupling constants, for which we assume ohmic spectral densities
\cite{Weiss99,LeggRev}
%
% NICK: la FORMULA e` CORRETTA perche` le \lambda hanno dimensioni di energia:
%
$J_{\xi}(\omega) = \sum_\nu \lambda_{\xi,\nu}^2 \delta(\omega-\omega_\nu)
 = \hbar^2\alpha_\xi \omega e^{-\omega/\omega_c}$.

Assuming the coupling to be weak, $\alpha_\xi \ll 1$, and retaining the
lowest-order in $\alpha_\xi$, the evolution of the system's reduced density
matrix $\tilde\rho_S$ in the rotating frame is given by the master equation
\cite{CohenT}
%
%\begin{widetext}
\begin{eqnarray}\label{ME}
&&\frac{\partial\tilde\rho_S(t)}{\partial t}
 \simeq -i [\Hef,\tilde\rho_S(t)]
  -\frac{1}{\hbar^2} \sum_{\xi=z,x} \int_0^{\infty} d\tau
\\ \nonumber
&&\left\{ G_\xi(\tau) \left[ \tilde\sigma^\xi(t),U_0^\dagger(-\tau)\tilde\sigma^\xi(t-\tau)U_0(-\tau)\tilde\rho_S(t) 
       \right] \right.
\\ \nonumber
&&+\left. G_\xi^*(\tau) \left[ \tilde\rho_S(t)
U_0^\dagger(-\tau)\tilde\sigma^\xi(t-\tau)U_0(-\tau),\tilde\sigma^\xi(t)
    \right] \right\}
,
\end{eqnarray}
%\end{widetext}
%
where $\tilde\sigma^\xi(t)= R_y^{-1}(\omega t) \sigma^\xi R_y(\omega t)$, and
$U_0(\tau)=\exp(-i\Hef\tau/\hbar)$.
The function $G_\xi(\tau)$ is expressed in terms of the spectral density as
\begin{equation}
G_\xi(\tau) = \int_0^\infty d\omega J_\xi(\omega)
\left[ \cos(\omega \tau) \coth \frac{\hbar\omega\beta}{2} - i\sin(\omega\tau) \right]
\;, \nonumber
\end{equation}
where $\beta=(k_{\rm B}T)^{-1}$, and $T$ is the bath temperature.

%%%%%%%%%%%%%%%%%%%%%%%%%%%%%%%%%%%%%%%%%%%%%%%%%%%%%%%%%%%%%%%%%%%%%%%%%%%%%%%
\begin{figure}\centering
\includegraphics[width=.45\textwidth]{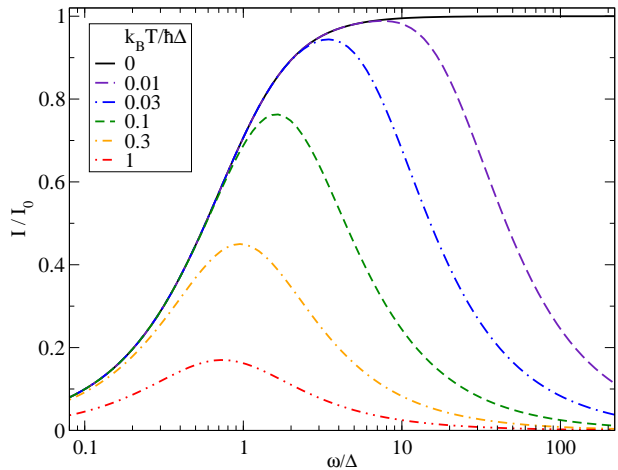}
\caption{\label{Temp:fig}
Steady-state DC circulating current $I$ as a function of the pumping
frequency $\omega$.
Solid line: $T=0$ current, obtained by Eq.~\eqref{current} with
$P\equiv1$.
Note that this  
%
% in fact I did it with T=10^-5, indistinguishable from 0 in this range!
%
coincides with the purely quantum result (no dissipation) when
$|\tilde{\psi}\rangle = |\hat{\bf n};- \rangle$
is chosen for the initial state.
Dashed and dot-dashed curves: pumped DC current for several temperatures,
as obtained by the exact expression \eqref{eq:main} for $P$.
}
\end{figure}
%%%%%%%%%%%%%%%%%%%%%%%%%%%%%%%%%%%%%%%%%%%%%%%%%%%%%%%%%%%%%%%%%%%%%%%%%%%%%%%

When the bath coupling to $\sigma_x$ and $\sigma_z$ have the same spectral
density ($J_x(\omega) = J_z(\omega) = J(\omega)$, even if non-ohmic) the
form of Eq.~\eqref{ME} becomes particularly simple, since all explicit time
dependence disappears and we are left with a constant-coefficients
inhomogeneous linear differential equation \cite{EPAPS_current}.
For $\alpha_\xi \rightarrow 0$ we find that the stationary density matrix
$\tilde\rho_S$ is diagonal in the basis $|{\bf n} \pm\rangle$, with a
polarization given by
\begin{equation} \label{eq:main}
P=\frac{(\omega'-\omega)^2 J_+ + (\omega'+\omega)^2 J_-}
       {(\omega'-\omega)^2 c_+J_+ + (\omega'+\omega)^2 c_-J_-} \;.
\end{equation}
Here $J_\pm = J(\omega'\pm \omega)$ and
$c_\pm=\coth[\hbar(\omega'\pm\omega)/(2k_{\rm B} T)]$.
The resulting DC circulating current, Eq.~\eqref{current}, is shown in
Fig.~\ref{Temp:fig} for a broad range of frequency and temperature.

Several  comments are in order.
(i) In the $T\to 0$ limit, irrespective of $\omega/\Delta$ and of the form
of the spectral density, the stationary master equation operator is a
projector onto the ground state $|{\bf n}-\rangle$ of $\Ham_{\rm eff}$.
%
% JOE: Tolto: fanno un altro modello ....
%This result extends the result of Ref.~\cite{Pekola10} to arbitrary
%frequency.
%
(ii) For $\omega\ll \Delta$, we have $P=\tanh[\hbar\Delta/(2k_{\rm B} T)]$
as is appropriate for a static Hamiltonian in thermal equilibrium.
At $T=0$, $P=1$ and the charge pumped in a period, $Q_p=2\pi I/\omega=2\pi
I_0/(\hbar\Delta)$ coincides exactly with the Berry-phase
result \cite{Pekola99} of Eq.~(21) in Ref.~\cite{MottonenPRB06}.
Nevertheless, the pumped charge, although geometric, is not quantized, and
depends on the ratio $\gamma_0/\Delta$.
It can be made quite large (though not arbitrarily large) by reducing the
amplitude of the perturbing field $\Delta$ in the range
$\hbar \omega \ll \hbar \Delta \ll \gamma_0$.
% IN CHE SENSO ESATTO E' GEOMETRICA CHIEDE PEKOLA, E C'ENTRA BERRY?(ERIO)
% HO SCRITTO BERRY ESPLICITAMENTE, ADESSO MI PARE ABBASTANZA CHIARO (NICK)
%
(iii) Finally, in the antiadiabatic regime $\omega \gg \Delta$  we find
$P=\tanh[\hbar(\omega'-\omega)/(2k_B T)]$.
This result is at first sight intriguing: for fast driving, the spin
reaches thermal equilibrium like a static spin Hamiltonian characterized by
an effective Larmor frequency
$(\omega'-\omega) = \Delta[\Delta/(2 \omega)+O(\Delta/ \omega)^3]$
that vanishes for large $\omega$.
The polarization, which remains identically unity for all large frequencies
at $T=0$, decays eventually 
at any finite $T$ for large $\omega$.
Faster and faster driving at finite temperature enhances the pumped current
up to $\omega\simeq \hbar\Delta^2/k_B T$.
For larger driving frequencies, thermal fluctuations catch up and suppress
$P$ causing the pumped current $I$ to drop, as seen on the high-frequency
side of Fig.~\ref{Temp:fig}.
The reason why $(\omega'-\omega)$ determines the Boltzmann occupancy of the
two levels split by $\hbar\omega'$ may be traced to the $-\omega\cdot {\bf
  M}$ term to be included in the thermodynamically relevant functions for a
body rotating at frequency $\omega$, see e.g., \textsection 26 of
Ref.~\cite{Landau5},
where ${\bf M}$ is the body angular momentum which, for our spin, coincides
with $\hbar \boldsymbol{\sigma}$.

The results just presented are analytical and thus quite elegant and
predictive.
Obtained as they were for a rather special case however, how general are
they?
To address this question we solve numerically Eq.~\eqref{ME} by means of
Runge-Kutta integration \cite{numrec}, and obtain
$\tilde\rho_S(t)$ and from that
$I(t)=I_0{\rm Tr} [\sigma^y \tilde\rho_S(t)]$.
The numerical approach allows us to study the effect of unequal
environments in the $x$ and $z$ directions, where the simplifications
leading to Eq.~\eqref{eq:main} do not hold.
In particular, we consider the case $\alpha_x\neq \alpha_z$ (by symmetry, it
does not matter which one is larger).
We find that at finite (but small) $\alpha_\xi$ the solution is no longer
stationary even in the rotating reference frame chosen, and small
oscillations of the density matrix and of the current at frequency
$2\omega$ remain undamped in the long-time limit.
Nevertheless for $\alpha_\xi \rightarrow 0$, the amplitude of these
oscillating density-matrix terms vanishes linearly with $\alpha_\xi$, and
the constant part of the density matrix at low temperature converges to the
symmetric-environment case.
In particular, at $T=0$ the polarization again saturates to 1.

This behavior for $\alpha_\xi\rightarrow 0$ can alternatively be recovered
by applying a rotating wave approximation to Eq.~\eqref{ME}, i.e., by
neglecting all the terms oscillating with frequency $\omega$ or $\omega'$.
Remarkably, the resulting equation again coincides without approximations
with the one appropriate to the symmetric environment.
We conclude that the results obtained for the symmetric environment are
indeed representative of those expected in the more general asymmetric
coupling case, provided the limit of weak coupling to the environment
holds.
In particular, Eq.~\eqref{eq:main} remains valid.

%%%%%%%%%%%%%%%%%%%%%%%%%%%%%%%%%%%%%%%%%%%%%%%%%%%%%%%%%%%%%%%%%%%%%%%%%%%%%%%
% NICK: inserita questa parte, mi pare illustrativa.
% Se non piace la si butta in EPAPS senza problemi
% NO, IO LA TERREI SE C'E' SPAZIO: IL TRANSIENTE E' INTERESSANTE, SE E'
% REALISTICO...(ERIO)
% NON SO QUANTO SIA REALISTICO, I PARAMETRI SONO SCELTI IN MODO CHE SI VEDA
% QUALCOSA, MA NON MATCHANO NESSUN CASO REALISTICO, PERCHE` NON SAPPIAMO
% STIMARE L'ACCOPPIAMENTO ALPHA IN UN CASO SPERIM. SIGNIFICATIVO.
% COMUNQUE LO SPAZIO ABBONDA! (NICK)

\begin{figure}\centering
\includegraphics[width=.45\textwidth]{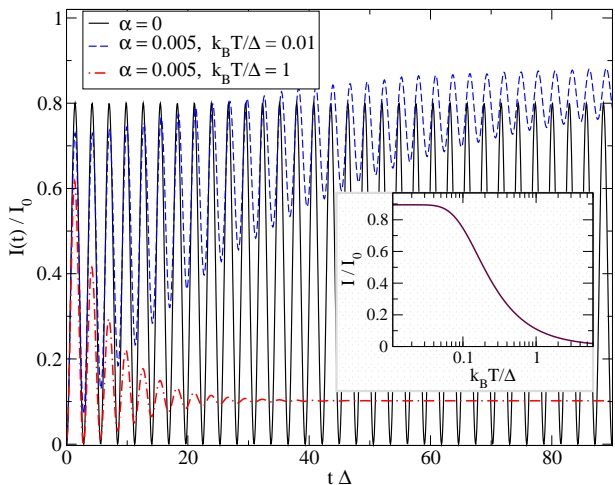}
\caption{\label{Evol:fig}
Time evolution of the current 
$I(t)=I_0{\rm Tr} [\sigma^y \tilde\rho_S(t)]$ for $\omega=2\Delta$,
in the dissipationless case ($\alpha=0$, solid curve), and in the
transient induced by weak dissipation ($\alpha=0.005$), at  low 
(dashed)  and intermediate temperature (dot-dashed), starting from
the initial $\ket{z,-}$ state.
Inset: temperature dependence of the steady-state dc current $I$.
}
\end{figure}
%%%%%%%%%%%%%%%%%%%%%%%%%%%%%%%%%%%%%%%%%%%%%%%%%%%%%%%%%%%%%%%%%%%%%%%%%%%%%%%

The numerical solution of the master equation \eqref{ME} also illustrates
the transient  approach to the stationary state.
Figure~\ref{Evol:fig} shows the full time evolution of the current,
compared to the pure quantum evolution in the absence of dissipation.
Note that, for a given coupling $\alpha$, temperature affects not only the
final steady current, via the final value of $P$, see Eq.~\eqref{eq:main} 
and inset of Fig.~\ref{Evol:fig}, but also the relaxation time with 
which this steady state is approached in the initial transient.

% CONTACT WITH PREVIOUS WORK:
In Ref.~\cite{Pekola10} an investigation was attempted of nonadiabaticity,
with numerical evidence that a stronger dissipation might somehow
compensate for the weak-coupling non-adiabatic current reduction relative
to the geometric value of the adiabatic limit.
Our exact solution clarifies that nonadiabaticity is fundamentally associated to 
such a radical current suppression that eventually, for large frequency, the charge pumped
in one period $Q_p\propto \omega^{-2}$.
The method introduced in a very recent work \cite{Russomanno10} also
can deal with the non-adiabatic
regime (indeed our $|{\bf n}\pm\rangle$ states
coincide with the Floquet states of that method), and represents a more
general approach to pumping problems in the presence of weak dissipation
\cite{Floquet:note}.
However in that work intermediate frequencies were studied only in the
absence of dissipation:
the present model seems unique in affording an explicit
analytic expression for the density matrix and current in the presence of
dissipation for arbitrarily high frequency.

\textit{Feasibility and quantitative estimates} --
Triple quantum dot systems have been recently realized experimentally
\cite{Gaudreau06}, and could be used to implement the pumping effect
proposed.
By adopting  a tentative hopping $\gamma_0\simeq 0.05$~meV
between the dots, we find a maximum current of the order of
$I_0 = 0.05~{\rm meV}\, q_{\rm e}/(\sqrt{3}\hbar) \simeq 7$~nA.
% = qE 0.00005 eV / Sqrt[3]/hbar
%
The magnetic field generated by this current could be detectable if the
dot-ring arrangement had at least an effective radius $r_\textrm{eff}\simeq
200$~nm \cite{Gaudreau09}.
In that case, a ring-shaped SQUID of $5\,\mu$m radius placed $\sim 5\,\mu$m
above the quantum dots would intercept a flux of order 0.02
%   \,\Phi_0$, where $\Phi_0= \pi \hbar/q_{\rm e}$ is the 
flux quanta, a value routinely detectable.

The frequency and temperature regions where this current enhancement could
be detected are determined by the frequency scale $\Delta$ of the effective
spin-$1/2$ model.
%
%The $\Delta$ value is the amplitude of the oscillating potentials acting on
%the dots, and is therefore under experimental control.
%
%However, the effective 2-level model is meaningful only for $\hbar \Delta \ll \gamma_0$, 
%or else one should include all three states in the calculation.
%
Assuming $\Delta \simeq 0.1 \gamma_0/\hbar \simeq 8$~GHz, the predicted
frequency-dependent dissipative effects on current should be observed near
and mainly above this resonant angular frequency at temperature $T\lesssim
0.2 \hbar\Delta/k_{\rm B}$, i.e., about $T\lesssim 0.01$~K for the
three-dot setup.
Our model could be relevant in a molecular context as well, where an
electronically degenerate point is looped about, as for example due to
cyclic molecular distortions, see Fig.~\ref{Model:fig} and
Ref.~\cite{EPAPS_current}.

In summary, we presented an analytical solution for the
time-dependent pumping of DC current in a quantum model with dissipation,
valid in the weak
% ohmic HO TOLTO, NON E` ESSENZIALE CHE SIA OHMICO, GIUSTO? (NICK)
dissipation limit.
The solution fully covers the crossover from the well-known adiabatic limit
to arbitrarily high frequencies.
The main physical surprise is that the frequency dependence of current
is nonmonotonic,
% and has a single peak.
%The peak position
with
% a maximum % AGGIUSTAMENTO
an optimal value
that moves from $\Delta$ upwards to infinity as temperature is reduced.
This effect, on the whole reminiscent of magnetic-resonance physics,
could be directly detectable for example in multi-dot arrangements.

We acknowledge helpful discussions with V. Brosco, R. Fazio, 
J.P. Pekola, and A. Russomanno.
Research was supported
by the Italian
% National Research Council
CNR through ESF Eurocore/FANAS/AFRI,
by the Italian Ministry of University and Research, through PRIN/COFIN 20087NX9Y7,
and by the French ANR, contract QNM ANR10-BLAN-0404-03.

%\bibliographystyle{unsrt}
%\bibliography{biblio}

\begin{thebibliography}{10}
\bibitem{Floquet:note}
  For the specific problem at hand, the rotating-wave approximation used in
  Ref.~\cite{Russomanno10} would lead to the same constant-coefficient
  equation as the one we derive from Eq.~\eqref{ME} without resorting to
  approximations.
\end{thebibliography}

\end{document}